\documentclass[a4paper,11pt]{article}
\usepackage{pos}
\usepackage{subfigure}
\usepackage{sidecap}
\usepackage[normalem]{ulem}

\begin{document}
\title{$J/\psi-$pair production at NLL in TMD factorisation at the LHC}

\let\OLDthebibliography\thebibliography
\renewcommand\thebibliography[1]{
  \OLDthebibliography{#1}
  \setlength{\parskip}{0pt}
  \setlength{\itemsep}{-1pt}
}

\author*[a]{Alice Colpani Serri}
\author[b,c]{Jelle Bor}
\author[b]{Dani\"el Boer}
\author[c]{Jean-Philippe Lansberg}

\affiliation[a]{Faculty of Physics, Warsaw University of Technology,\\plac Politechniki 1, 00-661, Warszawa, Poland}
\affiliation[b]{Van Swinderen Institute for Particle Physics and Gravity, University of Groningen,\\Nijenborgh 4, 9747 AG Groningen, The Netherlands}
\affiliation[c]{Universit\'e Paris-Saclay, CNRS, IJCLab, \\rue Georges Cl\'emenceau 15, 91405 Orsay, France}

\emailAdd{alice.colpani\_serri.dokt@pw.edu.pl}
\emailAdd{j.bor@rug.nl}
\emailAdd{d.boer@rug.nl}
\emailAdd{jean-philippe.lansberg@in2p3.fr}

\abstract{$J/\psi-$pair production at the LHC is currently one of the few tools available to probe gluon transverse momentum distributions (TMDs). In this context, data from LHCb in the collider mode have the potential to probe the evolution of the unpolarised-gluon TMDs and to measure the distribution of the linearly-polarised gluon in unpolarised protons for the first time.
In this proceedings contribution, improved predictions obtained for the LHC (at $\sqrt{s}$ = 13 TeV) up to next-to-leading logarithm (NLL) in TMD factorisation are presented.
We show the obtained predictions of transverse-momentum distributions at different invariant masses and rapidities computed in the LHCb acceptance along with PDF uncertainty. We predict the azimuthal modulations of the cross section that arise from linearly-polarised gluons.
}

\FullConference{The European Physical Society Conference on High Energy Physics (EPS-HEP2023)\\
 21-25 August 2023\\
Hamburg, Germany\\}

\maketitle

\vspace*{-0.3cm}
\section{Introduction}

Inclusive $J/\psi$-pair production in proton-proton collisions represents a great tool to allow for extractions of the poorly known gluon Transverse Momentum Dependent Parton Distribution Functions (TMD-PDFs or TMDs) \cite{Lansberg:2017dzg,Scarpa:2019fol}. Indeed, this process is mainly generated by gluon-gluon fusion and Color Singlet (CS) transitions are the main source of $J/\psi$ pairs, for which TMD-factorisation-breaking effects are absent \cite{Collins:2007nk,Collins:2007jp,Rogers:2010dm}.
For this reason $J/\psi$-pair production is considered a great candidate for probing gluon TMDs at the LHC. 
Moreover, the invariant mass of the $J/\psi$-pair in the final state can be tuned with the individual momenta of the two $J/\psi$, allowing for the investigation of the scale evolution of the TMDs.

\vspace*{-0.3cm}
\section{Overview of the process and formalism}

The process considered in our study is the simultaneous production of two $J/\psi$ in a single parton scattering from unpolarised proton-proton collisions. The $J/\psi$ is relatively easy to  produce and to detect, allowing for the collection of a large number of experimental data. From a theoretical point of view, though, it is still not clear how to treat quarkonium production: many models have been proposed in an attempt to describe quarkonium-production mechanisms. However, the consensus is that the \textit{Colour-Singlet Model} (CSM) works for the particular case where a $J/\psi$-pair is generated \cite{Lansberg:2019adr}.

\begin{figure}[b]
\centering
\includegraphics[width=10cm]{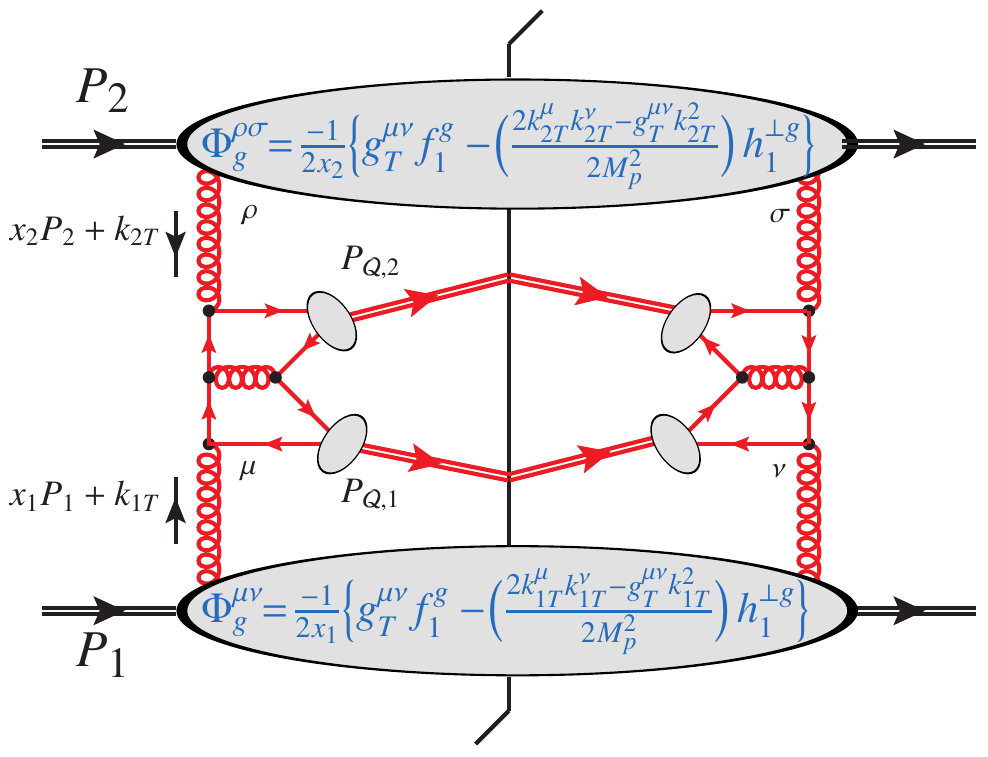}
\caption{Schematic overview of the inclusive scattering for $p+p\to J/\psi +J/\psi$ in TMD factorisation. From \cite{Scarpa:2020sdy}.}
\label{FigdoubleJpsi}
\end{figure}

A leading-order Feynman diagram of the process is shown in Figure \ref{FigdoubleJpsi}.
The protons have momentum $P_{1}$ and $P_{2}$ and the partons take a momentum fraction $x_{i}$ from them (collinear contribution), besides having a transverse component $k_{iT}$. Considering the TMD factorisation \cite{Collins:2011zzd}, the non-perturbative gluon TMDs are defined through the hadron correlator $\Phi(x_{i},k_{iT})$. For an unpolarised proton in particular, $\Phi(x_{i},k_{iT})$ is parameterised in terms of two independent TMDs at leading twist \cite{Mulders:2000sh}: the unpolarised gluon distribution, $f_{1}^{g}$, and the linearly-polarised gluon distribution, $h_{1}^{\perp g}$ (see Figure \ref{FigdoubleJpsi}).
The hadronic cross section for a gluon-fusion process, considering the production of a quarkonium pair, is given by the following equation \cite{Scarpa:2019fol}: 
\begin{align}
  \frac{d\sigma}{dM_{QQ}dy_{QQ}d^2q_Td\Omega}&=\frac{\sqrt{M_{QQ}^2-4M_Q^2}}{(2\pi)^28sM_{QQ}^2} \times \nonumber  
   \bigg \{F_1\mathcal{C}[f_1^gf_1^g]+F_2\mathcal{C}[w_2h_1^{\bot g}h_1^{\bot g}] \nonumber \\
&+\bigg{(} F_3\mathcal{C}[w_3f_1^{g}h_1^{\bot g}]+ 
F'_3\mathcal{C}[w'_3h_1^{\bot g}f_1^{g}] \bigg{)} \cos(2\phi_{CS}) \nonumber \\
&+F_4\mathcal{C}[w_4h_1^{\bot g}h_1^{\bot g}] \cos(4\phi_{CS}) \bigg{\}}, \label{HadrCross}
\end{align}
where $d\Omega=d\cos(\theta_{CS})d\phi_{CS}$, with $\theta_{CS}$ and $\phi_{CS}$ the polar and azimuthal Collins-Soper angles respectively.
$M_Q$ is the quarkonium mass (in our calculations for $\psi+\psi$, we took $M_Q=3.1\,\text{GeV}$), while $M_{QQ}$ indicates the invariant mass of the quarkonium pair, which we set as the hard scale of the process.
$y_{QQ}$ represents the rapidity of the quarkonium pair defined in the proton center-of-mass frame, i.e.:
\begin{align}
x_{1,2}=\frac{e^{\pm y}M_{QQ}}{\sqrt{s}}\label{Eqxs}\hspace{1mm},
\end{align}
with $s=(P_{1}+P_{2})^{2}$. The coefficients $F_i$ are the hard-scattering coefficients. They contain the explicit dependence on $M_{QQ}$ and the angle $\theta_{CS}$. The $\mathcal{C}[fg]$ are convolutions containing different combinations of $f_1^g$ and $h_1^{\perp g}$, in general:
\begin{align}
\mathcal{C}[wfg] = \int d^{2}k_{1T} \int d^{2}k_{2T} \hspace{1mm} \delta^{2}(k_{1T}+k_{2T}-q_T)\hspace{1mm} w(k_{1T},k_{2T})\hspace{1mm}f(x_{1},k_{1T}^{2})\hspace{1mm}g(x_{2},k_{2T}^{2})\hspace{1mm},
\label{EqConv}
\end{align}
where $k_{iT}$ are the transverse momenta of the gluons, $q_T$ is the transverse momentum of the quarkonium-pair and $w(k_{1T},k_{2T})$ are transverse weights \cite{Lansberg:2017tlc}. The azimuthal angle defined by the quarkonium pair in the final state is directly related to gluon TMDs \cite{Lansberg:2017dzg}:

\begin{align}
\langle\cos\hspace{1mm}(2\phi_{CS})\rangle&=\frac{1}{2}\frac{F_{3} \big(\mathcal{C}[w_{3} f_{1}^{g}h_{1}^{\perp g}]+\mathcal{C}[w_{3}' h_{1}^{\perp g}f_{1}^{g}]\big)}{F_{1} \hspace{1mm}\mathcal{C}[f_{1}^{g}f_{1}^{g}]+F_{2} \hspace{1mm} \mathcal{C}[w_{2} h_{1}^{\perp g} h_{1}^{\perp g}]} \hspace{1mm}, \label{EqMod1}\\
\langle\cos\hspace{1mm}(4\phi_{CS})\rangle&=\frac{1}{2}\frac{F_{4} \hspace{1mm}\mathcal{C}[w_{4} h_{1}^{\perp g}h_{1}^{\perp g}]}{F_{1} \hspace{1mm}\mathcal{C}[f_{1}^{g}f_{1}^{g}]+F_{2} \hspace{1mm} \mathcal{C}[w_{2} h_{1}^{\perp g} h_{1}^{\perp g}]}
\hspace{1mm}.
\label{EqMod2}
\end{align}
These are the expressions of the azimuthal modulations normalised to the azimuthally-independent part of the cross section. We note that such modulations are non-vanishing only if $h_{1}^{\perp g}$ is not zero.

\section{Switching on TMD evolution and results}
TMD-evolution studies are commonly implemented in impact-parameter space \cite{Collins:2011zzd}, $b_T$, in which the convolutions can be written as simple products \cite{Collins:1984kg}:
\begin{align}\label{eq:FFcfg}
{\cal C}[w\,f\,g](x_{1},x_{2},{q}_{T}; Q) 
 & = \int_{0}^{\infty} \frac{d b_{T}}{2 \pi} \, b_{T}^{n} \, J_{m}(b_{T}\,q_{T})   \, e^{-S_{A}(b_{T}^{*};Q^{2},Q)} \, e^{-S_{NP}(b_{T};Q)}\,  \nonumber \\
 & \times \,  \hat{f}(x_{1},{{b}}_{T}^{*};\mu_{b}^{2},\mu_{b}) \, \hat{g}(x_{2},{{b}}_{T}^{*};\mu_{b}^{2},\mu_{b}),
\end{align}
where $Q$ is the hard scale of the process (which we choose to be $M_{QQ}$) and $J_m(b_Tq_T)$ is the Bessel function of order $m$. $S_A$ is the perturbative Sudakov factor at next-to-leading-logarithmic (NLL) accuracy \cite{Collins:1981va, Bacchetta:2022awv} and $S_{NP}$ the non-perturbative one, chosen to be a Gaussian \cite{Scarpa:2019fol}:
\begin{align}
S_{NP}(b_{T};Q) =  A \, \text{ln} \frac{Q}{Q_{NP}}b_{T}^{2} \quad \text{with} \quad Q_{NP}=1\, \text{GeV}.
\label{Snp}
\end{align}
$\hat{f}$ and $\hat{g}$ are the Fourier-transformed gluon TMDs:
\begin{align}
\hat{f}_{1}^{g}(x,{{b}}_{T}^{2}) &\equiv \int d^{2}{{q}}_{T} \hspace{1mm} e^{i {{b}}_{T} \cdot {{q}}_{T}} \hspace{1mm}f_{1}^{g}(x,{{q}}_{T}^{2})\hspace{1mm}, \\
\hat{h}_{1}^{\perp g}(x,{{b}}_{T}^{2}) &\equiv \int d^{2}{{q}}_{T} \hspace{1mm} \frac{({{b}}_{T}\cdot{{q}}_{T})^{2} - \frac{1}{2} {{b}}_{T}^{2}{{q}}_{T}^{2}}{{{b}}_{T}^{2}M_{p}^{2}}\hspace{1mm}e^{i {{b}}_{T} \cdot {{q}}_{T}} \hspace{1mm}h_{1}^{\perp g}(x,{{q}}_{T}^{2}).
\end{align}
The expressions above are valid for $b_0/Q \leq b_T \leq b_{T,\text{max}}$ (with $b_{T,\text{max}}$ estimated to be around 1.5$\,\text{GeV}^{-1}$ and with $b_{0} = 2e^{-\gamma_{E}}$): when $\mu_b=\ b_{0}/b_{T}$ becomes larger than $Q$, the evolution should stop ($S_{A}=0$), while for values larger than $ b_{T,\text{max}}$ perturbation theory starts to become less reliable. 
To force the Fourier transform of these perturbative objects to remain in the range where they make sense, one  changes~\cite{Collins:2016hqq, Collins:1984kg} $b_T$ into two variants in Equation (\ref{eq:FFcfg}) as :
\begin{align}
b_{c}(b_{T}^{*})=\sqrt{(b_{T}^{*})^{2}+\bigg(\frac{b_{0}}{Q}\bigg)^{2}} \quad  \text{and} \quad
 b_{T}^{*}(b_{T}) = \frac{b_{T}}{\sqrt{1+\bigg({b_{T}}/{b_{T,\text{max}}}\bigg)^{2}}}.
\end{align}

Such a formalism has been implemented for quarkonium-pair production for the first time in \cite{Scarpa:2020sdy}, where TMD evolution effects have been shown to be measurable. We present updated results, taking into account also the $x$ and $y_{QQ}$ dependence and PDF (mstw2008lo \cite{Martin:2009iq}) uncertainty.
\begin{figure}[hb]
     \centering
     \begin{subfigure}
         \centering
         \includegraphics[width=7.45cm]{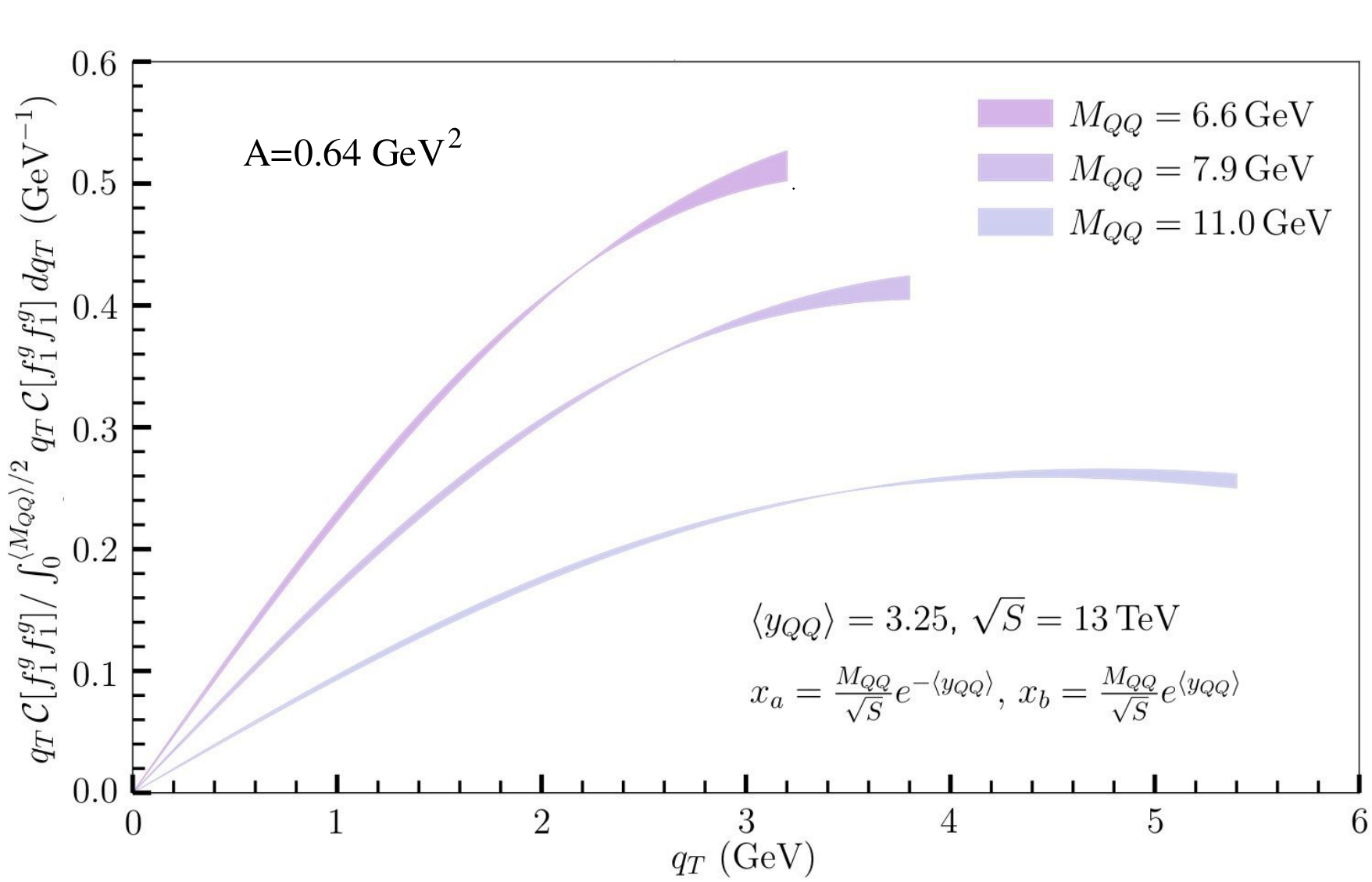}
     \end{subfigure}
     \begin{subfigure}
         \centering
          \includegraphics[width=7.45cm]{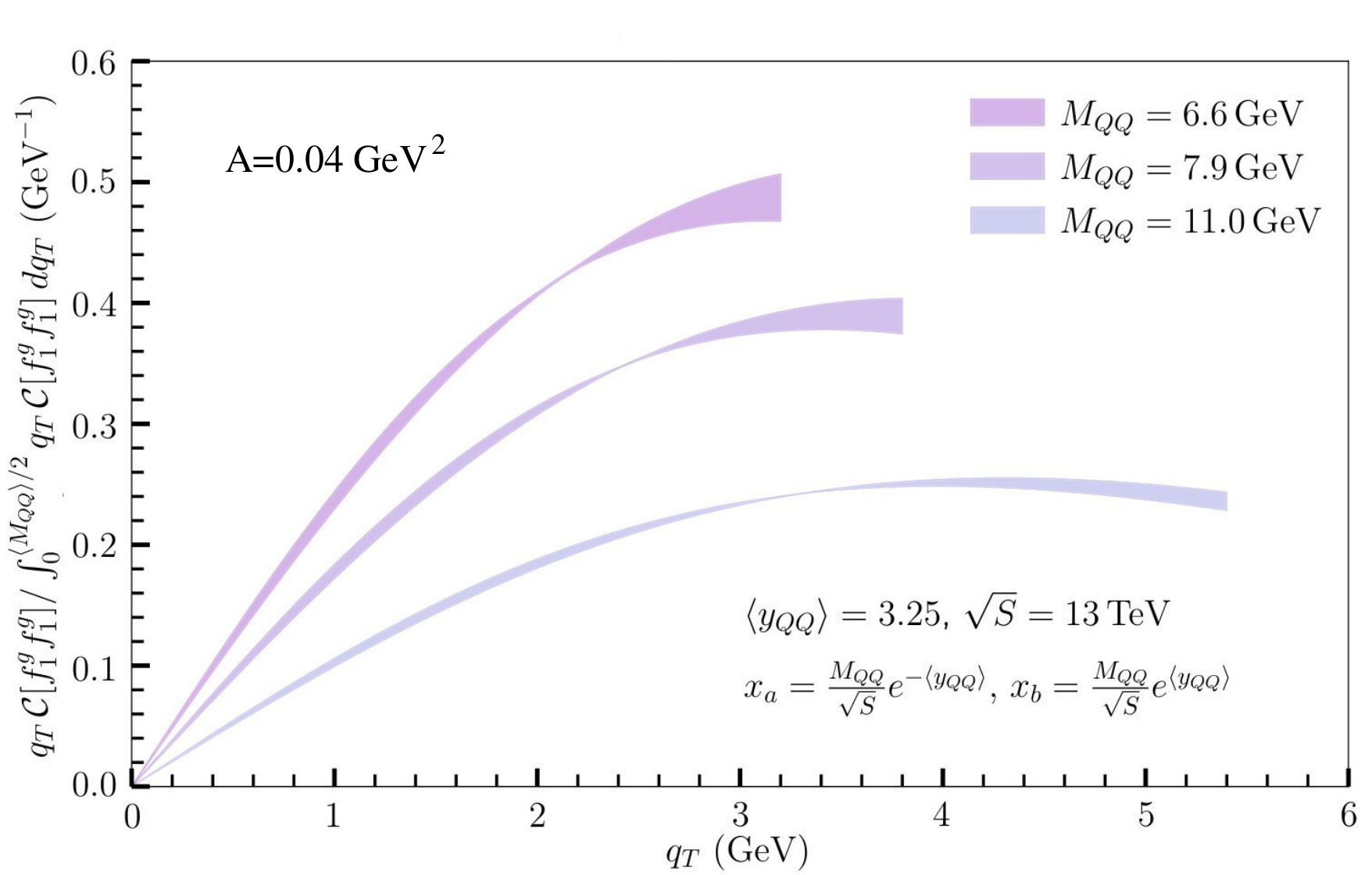}
         
     \end{subfigure}
    \vspace*{-0.4cm}
    \caption{Normalised $q_T$ spectrum at fixed rapidity ($y_{QQ}=$ 3.25) with three different values of the mass $M_{QQ}$ = 6.6, 7.9 and 11.0 GeV and $A$ = 0.64 GeV$^2$ (left) and 0.04 GeV$^2$ (right).    \vspace*{-0.1cm}}
\label{Norm_qT}
\end{figure}

The plots in Figure \ref{Norm_qT} show the normalised $q_T$ spectrum for $J/\psi$-pair production using the evolved TMDs at $M_{QQ}$ = 6.6, 7.9 and 11.0 GeV. These values are chosen according to the bins of the recent preliminary LHCb measurements \cite{Liupan} of this process. The width of each band represents the PDF uncertainty. The difference between left and right plots is given by the different chosen value of $A$ (Equation (\ref{Snp})): changing this quantity gives an estimate of the non-perturbative TMD uncertainty. The plots suggest that the peak is moving towards larger $q_T$ values when the scale increases, while the PDF uncertainty tends to decrease.
\begin{SCfigure}
    \centering
     \includegraphics[width=8cm]{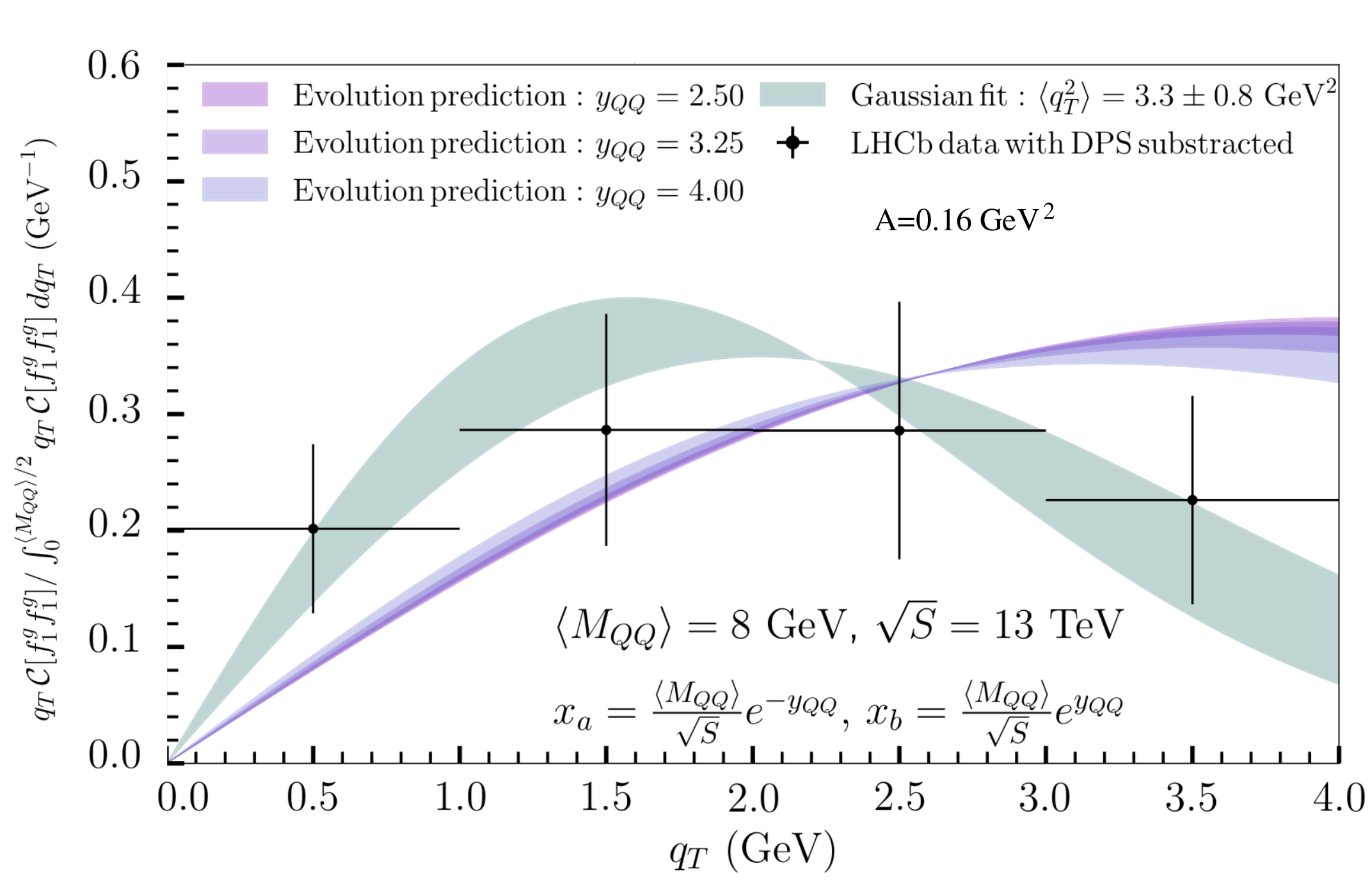}
    \caption{Normalised $q_T$ spectrum for 3 values of rapidity ($y$ = 2.50, 3.25, 4.00) at fixed $M_{QQ}$ (8 GeV) and $A$ = 0.16 GeV$^2$. The gray band represents the uncertainty of the gaussian fit made in \cite{Scarpa:2019fol} and the black crosses are the experimental data from LHCb published in 2017 \cite{LHCb:2016wuo} from which DPS had been subtracted \cite{Scarpa:2019fol}. \vspace*{-0.2cm}}
    \label{Norm_qT_andFit}
\end{SCfigure}
In Figure \ref{Norm_qT_andFit}, the $q_T$ spectrum is plotted with fixed invariant mass ($M_{QQ}$ = 8 GeV) and varying the rapidity ($y$ = 2.50, 3.25 and 4.00). Changing the rapidity slightly affects the $q_T$-spectrum behaviour. Again, the band widths correspond to the PDF uncertainty.

From Equations (\ref{EqMod1}) and (\ref{EqMod2}), we have computed the azimuthal modulations $\langle\cos\hspace{1mm}(2\phi_{CS})\rangle$ and $\langle\cos\hspace{1mm}(4\phi_{CS})\rangle$. We have performed a study in different $\cos{(\theta_{CS})}$ bins, namely [0, 0.25], [0.25, 0.50], [0.50, 0.75], [0.75, 1] and [0, 1] for $M_{QQ}$ = 8 and 11 GeV. Both modulations show a mass dependence at fixed $q_T$ and they become larger for a larger value of the mass (they increase of a factor $\sim 2$). We have found that $\langle\cos\hspace{1mm}(2\phi_{CS})\rangle$ is positive in all bins and that it increases while increasing $q_T$ and has higher contribution for $\cos{(\theta_{CS})} \in$ [0.50, 0.75] (expectation of $\mathcal{O}(\text{few}\%)$).
$\langle\cos\hspace{1mm}(4\phi_{CS})\rangle$ has higher positive contribution for $\cos{(\theta_{CS})} \in$ [0, 0.25], then it becomes negative at $\cos{(\theta_{CS})} \sim 0.3$ after which it reaches the highest contribution (in absolute value) for $\cos{(\theta_{CS})} \in$ [0.50, 0.75] (expectation of $\mathcal{O}(1\%)$). In both cases, we have large non-perturbative uncertainties. Such preliminary results are compatible with the latest experimental data from LHCb \cite{Liupan}, presented at this conference, from which $\langle$cos$(2\phi)\rangle=-0.029\pm0.050\pm0.009$ and $\langle$cos$(4\phi)\rangle=-0.087\pm0.052\pm0.013$ have been found considering the overall $\cos{(\theta_{CS})}$ region.

\vspace*{-0.3cm}
\section{Conclusions and outlook}
 \vspace*{-0.1cm}
Quarkonium production has the potential to probe the internal structure of the nucleon. Double-$J/\psi$ production in particular gives the possibility to investigate gluon TMD-induced effects and $\Upsilon$ predictions could be studied soon. TMD evolution effects are measurable and we have shown the obtained predictions considering $x$ and rapidity dependencies and PDF uncertainty for the first time. 

We are currently working on a novel method to determine the non-perturbative Sudakov factor because certain issues with a simple Gaussian Ansatz have been identified, which improve our predictions for double-$J/\psi$ production.
In the future, studies could be performed considering polarised protons (like in the fixed-target experiments at the LHC) in order to access more gluon TMDs. In particular, double-$J/\psi$ production is the most promising process for accessing the gluon Sivers function \cite{HADJIDAKIS20211}.

\vspace*{-0.2cm}

\section*{Acknowledgements}\vspace*{-0.35cm}
This work was supported in part by the European Union’s Horizon 2020 research and innovation program under Grant Agreements No. 824093 (Strong2020) in order to contribute to the EU Virtual Access ``NLOAccess''. This project has also received funding from the French Agence Nationale de la Recherche (ANR) via the grant ANR-20-CE31-0015 (``PrecisOnium'')  and was also partly supported by the French CNRS via the COPIN-IN2P3 bilateral agreement. 

\vspace*{-0.35cm}
\bibliographystyle{JHEP}
\bibliography{biblio}

\end{document}